\documentclass[preprint2,longabstract]{aastex}

\shorttitle{ORM site characteristics}
\shortauthors{Lombardi et al.}

\begin{document}

\title{El Roque de Los Muchachos site characteristics, \\
    I.Temperature analysis}

\author{G. Lombardi$^{a, b}$, V. Zitelli$^{b}$, S. Ortolani$^{c}$ and M. Pedani$^{d}$}
\affil{}
\affil{\small\upshape\rmfamily{($a$) Astronomy Department, University of Bologna, via Ranzani 1, I-40127 Bologna, Italy}}
\affil{\small\upshape\rmfamily{($b$) INAF $-$ Bologna Astronomical Observatory, via Ranzani 1, I-40127 Bologna,
Italy}}
\affil{\small\upshape\rmfamily{($c$) Astronomy Department, University of Padova, vicolo dell'Osservatorio 2, I-35122 Padova, Italy}}
\affil{\small\upshape\rmfamily{($d$) Fundaci\'on Galileo Galilei and Telescopio Nazionale Galileo, PO Box 565, E-38700, La Palma, Tenerife, Spain}}
\affil{}
\affil{\small\rmfamily{Send offprint requests to:} \small\upshape\ttfamily{gianluca.lombardi@oabo.inaf.it}}

\begin{abstract}
  In this paper we present an analysis of temperature taken at two telescopes located at the
   Observatorio del Roque de Los Muchachos in the Canary Islands. More than 20 years of measurements
    at CAMC are included.
   The analysis of the data from TNG and CAMC are compared in order to check local variations and long term trends.
   Furthermore, the temperatures at different heights are correlated to the quality of astronomical seeing.
   We considered the correlation of NAO Index and annual down$-$time with mean annual temperatures.
   The final aim of this work is to better understand the influence of wide scale parameters on local
    meteorological data.
   The analysis is done using a statistical approach. From each long series of data we compute the
    hourly averages and than the monthly averages in order to reduce the short$-$time fluctuations due
    to the day/night cycle. A particular care is used to minimize any effect due to biases in case of lacking
     of data. Finally, we compute the annual average from the monthly ones.
   The two telescopes show similar trends. There is an increase of temperatures of about
    1.0$^{\circ}$C/10yrs from the annual means and a more rapid increase of the annual minimums
    then the maximums. We found that positive NAO Index reduces the increase of temperatures,
      and accelerates the decrease. Moreover, there is no evidence that positive NAO Index corresponds to a lower number of
      non-observable nights. Finally, seeing deteriorates when the gradient of temperatures between
      2 and 10 m above the ground is greater than $-0.6^{\circ}$C.
\end{abstract}

\keywords{site testing}

\section{Introduction}
Since the year 1970, La Palma Island, located at about 400 km off the Moroccan coast of North$-$West Africa, 
appeared a favorable geophysical site from the point of view of the sky conditions, due to the proximity of 
the semi-permanent Azores high pressure system, and it was chosen to host the main astronomical telescopes. 
It is known that the very good astronomical conditions of the island are mainly due to the stable subsiding 
maritime airmass that place most of the time the telescopes near the top of the mountain well above the 
inversion layer occurring in the range between 800 m and 1200 m (McInnes \& Walker \cite{mcinnes}).\\
All the telescopes are located along the northern edge of the Caldera de Taburiente, 
at the N$-$W side of La Palma Island, where the irregular shapes produce a complex orography
and the crowdedness of the top, due to the presence of all the astronomical observatories, suggests 
a possible modification of the local microclimate making difficult to foresee in advance the
precise local meteorological parameters. Therefore in these last years the ORM has been extensively monitored 
thanks to the efforts of the several site testing groups belonging to the hosted
astronomical observatories.\\
Since several years, various astronomical sites are monitored on a continuous basis by automatic weather
 stations, which provide measurements of a few local meteorological parameters. All these
instantaneous and long term records of the meteorological data are important tools for meteorological 
and climatological studies, as well as for the calibration of satellite remote sensing of the atmospheric 
and ground conditions (Zitelli et al. \cite{zitelli}).\\
In this paper we present for the first time an analysis of measurements obtained from local meteorological
towers and environmental conditions made at three telescopes at ORM.
The meteo data of Telescopio Nazionale Galileo (TNG) and Carlsberg Automatic Meridian Circle Telescope
(CAMC) are compared in order to check local variations or meteorological
 conditions. The analysis of these differences in terms of image quality at the telescopes will be discussed
  in this paper as follows:\\
In \S 2 we discuss the annual temperature means for the two telescopes, in \S 3
we present differences between day$-$time and night$-$time mean values and their comparison
with the down$-$time at CAMC, in \S 4 it is presented the North Atlantic Oscillation Index and its correlation with temperatures, in \S 5 we present seeing and temperature analysis for TNG. A preliminary comparison among the 
three sites \S 6 it is also presented on the basis of the results of the previous sections.

\section{Annual data analysis}\label{ann-data}
In this section we describe air temperatures (T) obtained by an accurate analysis of the meteorological data
from TNG and CAMC meteo stations which are distant of about 1000 m. The two sites are located well above the
inversion layer as shown in Table \ref{telescopes} that lists positions and heights of the telescopes.
   \begin{table}[h!]
     \begin{center}\small
       \caption[]{Locations and heights of TNG and CAMC.}
    \label{telescopes}
        \begin{tabular}{l | c | c | c }
           \hline
            \noalign{\smallskip}
            & Lat. [N] & Long. [W] & Height [m]\\
            \noalign{\smallskip}
            \hline
            \noalign{\smallskip}
TNG & 28$^{\circ}$45'28.3''& 17$^{\circ}$53'37.9''& 2387\\
CAMC & 28$^{\circ}$45'36.0''& 17$^{\circ}$52'57.0''& 2326\\
            \noalign{\smallskip}
            \hline
         \end{tabular}
         \end{center}
   \end{table}
The TNG meteo tower is a robust steel structure with a total height of 15 m. The tower is located about
 100 m far from TNG building. The temperature sensors are distributed along the tower at different
heights (ground, 2, 5 and 10 m) and the data are regularly sent from the tower to TNG annex building by 
means of an optic fiber link since 27 March 1998. The data sampling rate is 10 seconds, while data storage
 is done every 30 seconds (Porceddu et al. \cite{porceddu},  Zitelli et al. \cite{zitelli}).\\
The CAMC carried out regular meteorological observations from a sensor at 10.5 m on the meteo tower, in 
the period 13 May 1984 to 31 March 2005 and the records are more or less continuous in that
period. For the years 1984, 1985 and 1986 meteorological readings are only available at 30 minute intervals. 
From January 1987 readings were made at 5 minute intervals throughout the day and
night regardless of whether observing was in progress. Beginning in December 1994, all readings were made 
at 20 seconds intervals and then averaged over 5 minutes\footnote{http://www.ast.cam.ac.uk}.\\
\begin{table}[t!]
     \begin{center}\small
       \caption[]{TNG and CAMC annual T averagers [$^{\circ}$C].}
    \label{annual-t-avrg}
       \begin{tabular}{l | c | c}
	 \tableline\tableline
            \noalign{\smallskip}
            & TNG & CAMC\\
            \noalign{\smallskip}
            \hline
            Year & T$_{10}$ & T\\
\tableline\tableline
1985 &   $-$    &   8.8\\
1986 &   $-$    &   8.9\\
1987 &   $-$    &   9.1\\
1988 &   $-$    &   7.4\\
1989 &   $-$    &   5.2\\
1990 &   $-$    &   8.8\\
1991 &   $-$    &   8.7\\
1992 &   $-$    &   7.9\\
1993 &   $-$    &   7.0\\
1994 &   $-$    &   9.8\\
1995 &   $-$    &   9.5\\
1996 &   $-$    &   8.6\\
1997 &   $-$    &   8.9\\
1998 &  10.1    &  10.0\\
1999 &   9.6    &   9.3\\
2000 &   9.9    &   9.6\\
2001 &  10.7    &  10.1\\
2002 &   9.7    &   9.6\\
2003 &   9.7    &   9.8\\
2004 &   8.9    &   9.0\\
2005 &   9.5    &   $-$\\
\tableline\tableline
         \end{tabular}
         \end{center}
\end{table}
Each meteo sensor provides an accuracy of 0.1$^{\circ}$C and the data series have to be considered in local time. From each raw data series of T we compute the hourly
 averages and then from each set of them we compute the monthly averages. This is useful to reduce the short$-$time
 fluctuations due to natural day/night cycle.\\
A particular care was used to minimize any effect due to biases in case of lacking of data that typically occurred 
in winter time. For each missing month value we take into account the average obtained
from the two  corresponding months in other years in which the values of the months before and after the absent one
 are similar. 
For example, if the lacking month is September 2002, we look for the two Augusts and Octobers in the other years 
having similar mean values of August and October 2002. The accepted September 2002 value is the  average of the 
Septembers corresponding to the chosen Augusts and Octobers. This is the main reason why we decided to use 
monthly averages as an intermediate step in the calculation of the annual averages. Finally, we computed the annual averages of T from the monthly ones for the three telescopes.\\
Because the TNG four temperature sensors are located at different heights, we carefully inspected the single values.
 We found that the sensors at 2, 5 and 10 m show practically the same results, but the
ground sensor gives a systematically higher temperatures. In the present study we chose the 10 m temperature
 (T$_{10}$) that is also the one corresponding to the height of the primary mirror inside the TNG dome.\\
Table \ref{annual-t-avrg} reports the annual mean computed values. We measured a samples of instantaneous 
temperatures differences of simultaneous data from CAMC and TNG, in the months of March 2004,
August 2004, November  2004 and January 2005, in order to estimate an upper limit in the errors of the computed 
annual values.
We found a RMS range between 0.1 and 0.5$^{\circ}$C. Comparing the instrumental accuracy of 0.1$^{\circ}$C
we see a good agreement between the two instruments and we can assume that the residual difference is due to
atmospherical local variation.\\
   \begin{figure}[t!]
   \centering
   \includegraphics[width=8cm]{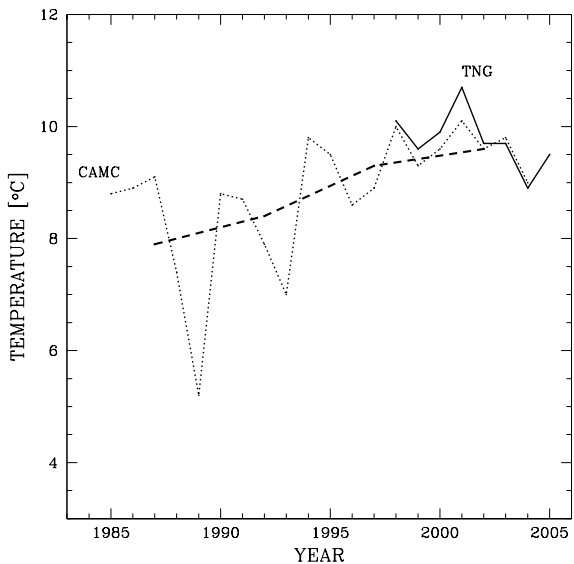}
   \caption{Annual temperatures at TNG (solid line) and CAMC (dotted line). The short-dashed
   line indicates the five$-$year average calculated on CAMC data series.}
              \label{t-comparison}
    \end{figure}
   \begin{figure}[t]
   \centering
   \includegraphics[width=8cm]{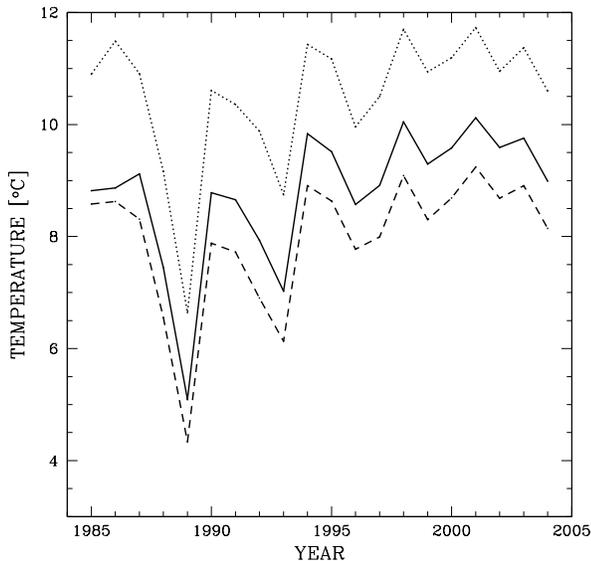}
   \caption{CAMC annual day$-$time (dotted line) and night$-$time (dashed line) temperature variations in comparison
    with entire day (solid line).}
              \label{test-day-night}
    \end{figure}
In Figure \ref{t-comparison} we show the plot of the annual values reported in Table \ref{annual-t-avrg}.
An increasing trend over the 20 years baseline of CAMC appears from the data. The best linear fit of those data
gives an increase of the temperatures of about 1.0$^{\circ}$C/10yrs.\\
The year 1989 appears to be the coldest one in our CAMC sample and 2001 is the warmest for all telescope sites.
 The data from the CAMC and TNG are remarkably similar, with average temperatures differing no more than 0.6$^{\circ}$C (year 2001).\\
We note an oscillation of the values with a period of about 3$-$4 years that seems to be slightly smoothed during
the last 10 years. Another evidence is that in the oscillation the points of local minimum and local maximum have
 a different behaviour, in fact the minimums increase more rapidly then the maximums.\\
Further tests about our \textit{modus operandi}, in order to check the reliability of the results have been done. In the first we computed the annual averages using direct raw data (bypassing hourly and monthly averages) gives
 almost identical mean values.\\
In the second test, the annual averages are derived from day$-$time (10$-$16 hrs) and night-time (22$-$4 hrs) 
data following the recipe of Jabiri et al. \cite{jabiri}. The results are reported in Figure
\ref{test-day-night} and  confirm the trend of Figure \ref{t-comparison}.\\
\begin{figure}[t]
   \centering
   \includegraphics[width=8cm]{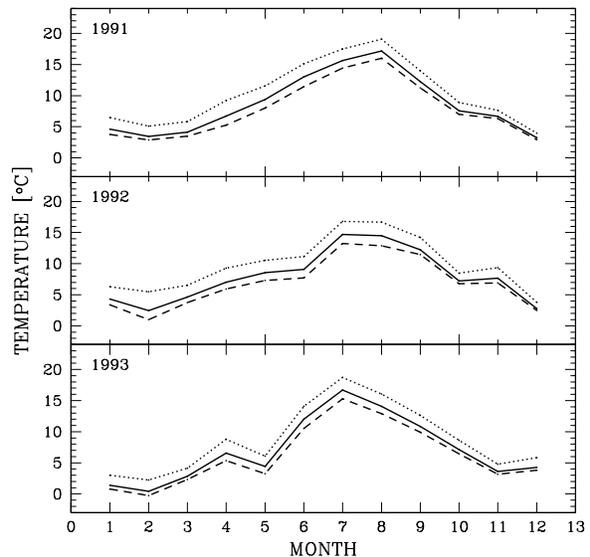}
   \caption{CAMC $-$ Monthly variation of entire day (solid line), day$-$time (dotted line) and night$-$time temperatures (dashed line) for 1991, 1992 and 1993.}
              \label{test-day-night-monthly}
    \end{figure}
\begin{figure}[t]
   \centering
   \includegraphics[width=8cm]{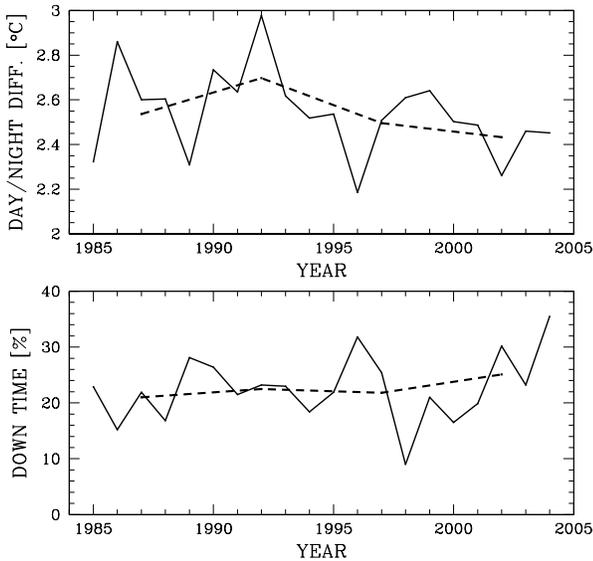}
   \caption{CAMC $-$ Trend of the differences between annual averages of day$-$time and night$-$time temperatures (top). Annual percentages of down$-$time due to weather (bottom). The short-dashed lines indicate the five$-$year average.}
              \label{day-night-downtime}
    \end{figure}
\begin{figure}[t]
   \centering
   \includegraphics[width=8cm]{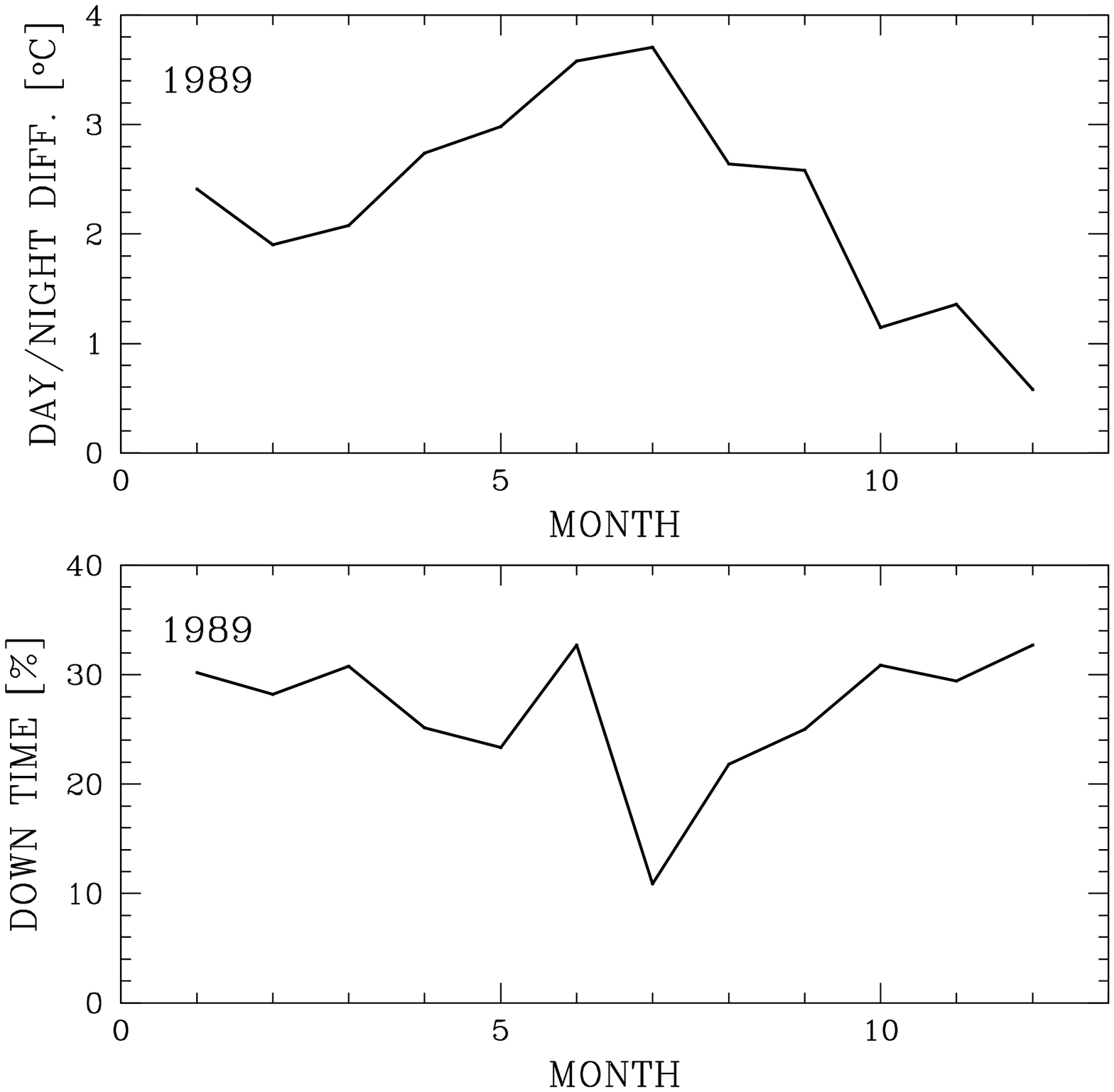}
   \caption{CAMC, year 1989 $-$ Differences between monthly averages of day$-$time and night$-$time temperature
   compared with monthly percentages of down$-$time due to weather.}
              \label{day-night-downtime89}
    \end{figure}
\begin{figure}[t]
   \centering
   \includegraphics[width=8cm]{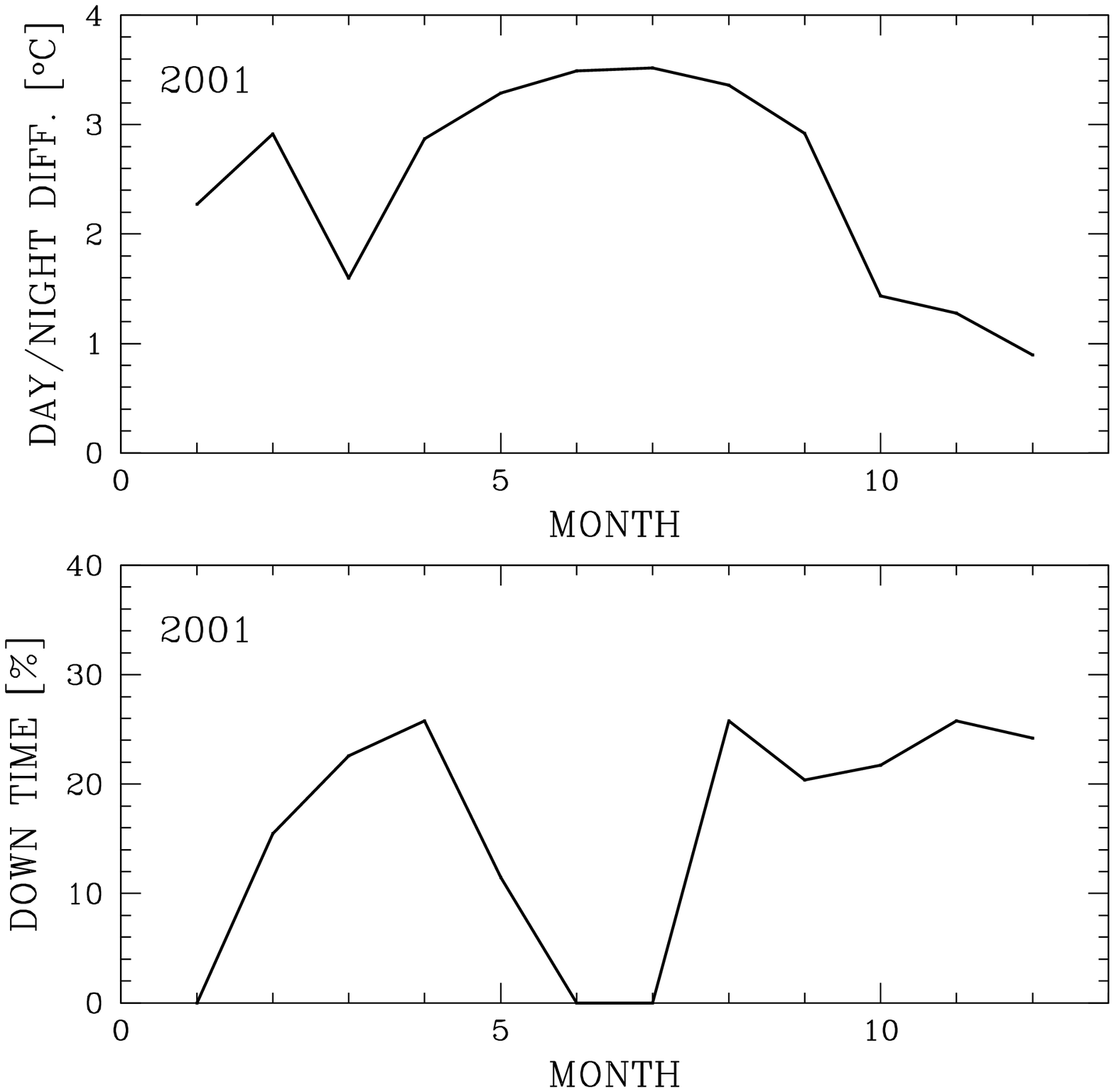}
   \caption{CAMC, year 2001 $-$ Differences between monthly averages of day$-$time and night$-$time temperature
   compared with monthly percentages of down$-$time due to weather.}
              \label{day-night-downtime01}
    \end{figure}
It is interesting to note that in Figure \ref{test-day-night} the mean annual temperatures are closer to the night$-$time temperatures then the day$-$time ones.
Using our monthly means we have carefully investigated if this is a result of a seasonal bias, but no evidence
 has been found (examples for 1991, 1992 and 1993 are reported in Figure \ref{test-day-night-monthly}). Jabiri et al. \cite{jabiri} show a very similar trend for monthly analysis of temperatures at CAMC in the period 1990$-$1993.\\

\section{Day$-$time and night$-$time variation}\label{day-night}
The annual averages of the differences between day$-$time and night$-$time temperatures ($\Delta$T) have
 been computed and the results are reported in Figure \ref{day-night-downtime} (top).
Also in this plot, the oscillations of the $\Delta$T seem to reduce the amplitude during the years.\\
Comparing the trends of annual temperatures of CAMC (Figure \ref{t-comparison}) and the corresponding $\Delta$T
(Figure \ref{day-night-downtime}, top), we note some correlation between the two trends because coolest years
present a small value of difference in temperature from day to night and vice versa. A Spearman correlation test between the two data series gives a significance of 50\%, very far from what we espected.\\
It is known that the decreasing diurnal temperature range would be linked to an increase in cloud coverage, generating a faster rate in the increase of daily minimums than in the maximums. We conclude that the effect can be explained as the direct influence of the number of cloudy days during the year that doesn't permit the sun to warm the atmosphere. This explains the correlation between the lost nights at CAMC due only to weather conditions (down$-$time), and $\Delta$T. As shown in Figure \ref{day-night-downtime} (bottom), high values of $\Delta$T correspond to lower down$-$time. The Spearman correlation test in this case gives a significance of 97\%.\\
To better see this difference we investigate two particular years: 1989 (the coolest one) and 2001 (the warmest).
In Figures \ref{day-night-downtime89} and \ref{day-night-downtime01} the behaviour is confirmed, with the exception
of some peculiar cases of some months (in particular in summer time) that shall be
investigated taking into account other parameters like pressure, wind direction and Sahara dusts (paper in preparation).

\section{North Atlantic Oscillation analysis}\label{NAO}
The North Atlantic Oscillation (NAO) is the dominant mode of atmospheric circulation in North Atlantic region
(Wanner et al. \cite{wanner}). It consists of a north$-$south dipole of pressure anomalies, with one center located
over Greenland and the other center of opposite sign spanning the central latitudes of the North Atlantic between
 35$^{\circ}$N and 40$^{\circ}$N.\\
NAO Index is generally defined as the difference in pressure between the Azores High and the Icelandic Low. The
positive phase of the NAO reflects below$-$normal heights pressure across the high latitudes of the North Atlantic
 and above$-$normal heights pressure over the central North Atlantic, the eastern United States and western Europe.
 The negative phase reflects an opposite pattern of pressure anomalies over these regions. The NAO exhibits considerable
  interseasonal and interannual variability, and prolonged periods (several months) of both positive and negative
   phases of the pattern are common\footnote{http://www.cpc.ncep.noaa.gov}.\\
Because the great influences of the NAO on the meteo conditions in
the northern emisphere, it is important to investigate a possible
correlation between NAO and other key parameters determining good or
bad observing astronomical conditions.
 For these reasons we calculated the annual averages for NAO from the monthly ones retrieved from the National Weather
 Service web site.
\begin{figure}[t]
   \centering
   \includegraphics[width=8cm]{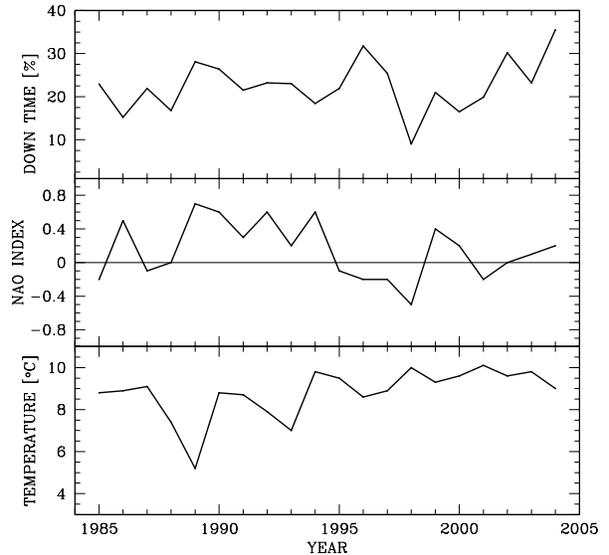}
   \caption{(center) North Atlantic Oscillation Index vs: (top) down$-$time due to weather conditions at the CAMC and (bottom)
   annual temperatures at CAMC site.}
              \label{t-VS-NAO}
    \end{figure}
In Figure \ref{t-VS-NAO} (top and center) the NAO Index from 1985 to 2004 and the respective down$-$time at CAMC due to weather
 conditions is compared. No correlation is found between NAO Index and number of non-observing nights (the significance in this case is about 25\%). A carefully inspection shows some particular effects of delay or different correlations (i.e. years 1989 and 1998) probably caused by peculiar events that should be investigated in other studies.\\
Figure \ref{t-VS-NAO} also shows the comparison between NAO Index (center) and
temperatures (bottom) computed from CAMC data archive. In this case, the
correlation between the amount of variability of the temperatures
year by year and the trend of the respective NAO Index has 86\% of significance. The
action of positive NAO Index is like a brake for the increase of
temperatures, and like an accelerator for the decrease. Vice versa, negative NAO Index acts in opposite mode.\\
An interesting point is the assumption of Graham \cite{graham2} that there is a poor correlation on annual basis
between NAO and air temperature from Mazo Airport in La Palma. This seems to give an indication that the influence of
the NAO above or below the inversion layer is different.\\
\begin{figure}[t]
   \centering
   \includegraphics[width=8cm]{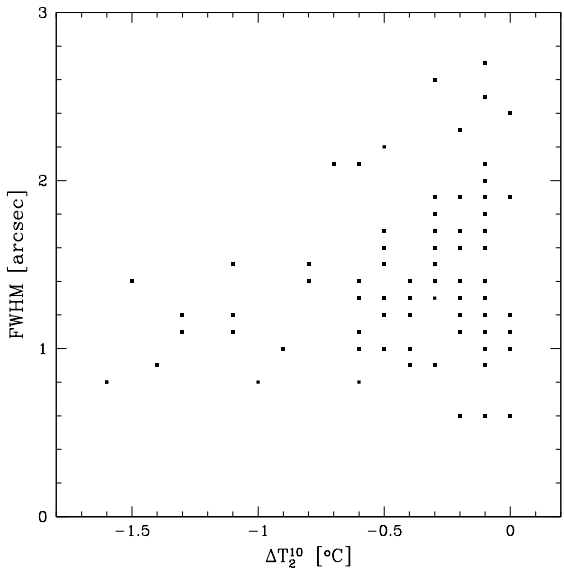}
v   \caption{Seeing in V band VS $\Delta$T$^{10}_{2}$ at TNG.}
              \label{DT-FWHM}
    \end{figure}

\section{Seeing and temperatures}\label{seeing}
We also investigated the influence of the temperature on the
astronomical seeing. Making use of 118 images obtained at TNG, pointed at zenith, from
31 January to 4 February 2000 with the image camera OIG, we computed
FWHM of several  stellar images in the V band frames. The images have been processed following the standard procedure (bias subtraction and flat fielding) using IRAF packages.\\
Following Racine et al. \cite{racine}, we check if any correlation exists
between the temperature T$_{M1}$ of the TNG primary mirror (M1) and T$_{10}$. The comparison between the monthly averages shows no appreciable difference, so we can consider T$_{10}$ as the temperature of M1.\\
We compute the gradients of temperature as $\Delta$T$^{10}_{2}$ = T$_{2}$ $-$ T$_{10}$ (where T$_{2}$ is the temperature
 measured by the sensor at 2 m) at the same UTs of the 118 images of which we computed the FWHM.\\
Figure \ref{DT-FWHM} displays the comparison between $\Delta$T$^{10}_{2}$ and the FWHM. In the plot the seeing
deteriorates when $\Delta$T$^{10}_{2}$ $> -0.6^{\circ}$C. This can be explained as consequence of the lower temperature
at 2 m because the higher temperature at 10 m inhibits the thermal convection below the primary mirror height.
\section{Cloudness and sites comparison}\label{site}
The fraction of available telescope time is one of the highest requirements to select astronomical sites, in particular
 night$-$time cloudness is strongly correlated with the closing of the dome.\\
\begin{figure}[t]
   \centering
   \includegraphics[width=8cm]{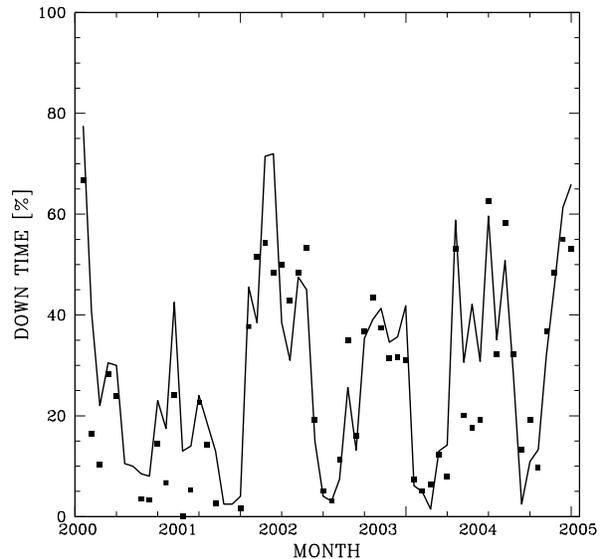}
   \caption{Comparison between monthly down$-$time from 2000 to 2004 for TNG (solid line) and CAMC (dots).}
              \label{monthly-downtime}
    \end{figure}
\begin{table}[t]
     \begin{center}\small
       \caption[]{TNG and CAMC percentages of annual down$-$time due to weather.}
    \label{2downtime}
         \begin{tabular}{l | c c c c c}
            \hline
            \noalign{\smallskip}
                      & 2000 & 2001 & 2002 & 2003 & 2004\\
            \noalign{\smallskip}
            \hline
            \noalign{\smallskip}
                       TNG & 27.4 & 27.6 & 26.1 & 28.2 & 37.3\\
                       CAMC & 16.5 & 19.9 & 30.2 & 23.2 & 35.5\\
            \noalign{\smallskip}
            \hline
         \end{tabular}
         \end{center}
\end{table}
Now there are long term records from many telescopes, with the number of available nights $-$ in particular the
CAMC records $-$ starting in 1984, that are the longest available at La Palma, covering a baseline of 22 years.
Records from the TNG report data starting from the year 2000 up to now\footnote{http://www.tng.iac.es}.\\
All these data are in excellent agreement. In particular the
comparison between weather condition of TNG and CAMC from 2000
to 2004 shows no evidence for a systematic trend of the weather time lost, but a seasonal periodic oscillation appears.\\
Table \ref{2downtime} reports the yearly mean values of the down$-$time weather, while Figure \ref{monthly-downtime}
reports the plot of the monthly averages.\\
The first detailed reports on the night$-$time cloudiness at La Palma are from Murdin \cite{murdin} that
reported 78\% of the usable nights in La Palma during the period of February$-$September 1975 (see Table 2
in Murdin \cite{murdin}). Restricting our set of data to the same range of Murdin (Februray$-$September) we found new percentages of down$-$time as reported in Table \ref{downtime-murdin}.\\
\begin{table}[t]
     \begin{center}\small
       \caption[]{TNG and CAMC complessive percentages of down$-$time due to weather, restricted in periods February$-$September.}
    \label{downtime-murdin}
         \begin{tabular}{l | c c c c c}
            \hline
            \noalign{\smallskip}
                       & 2000 & 2001 & 2002 & 2003 & 2004\\
            \noalign{\smallskip}
            \hline
            \noalign{\smallskip}
                       TNG & 12.7 & 10.5 & 15.3 & 13.2 & 19.9\\
                       CAMC & 7.0 & 7.0 & 17.9 & 11.6 & 21.1\\
            \noalign{\smallskip}
            \hline
         \end{tabular}
         \end{center}
\end{table}
The two telescopes, in spite of their distance of  about 1000 m and difference in height of about 60 m (cfr. Table 1), seem
to have a marginal different amount of down$-$time.\\
There is evidence that in the last years the fraction of lost observing time at CAMC and TNG is increasing.
On average, the annual values reported in Table \ref{2downtime} are considerable higher than the 22\% estimated by Murdin \cite{murdin}. Instead, the restricted data show lower number of lost nights. This may reflect that we are in presence of a strong seasonal effect.\\
To fully understand this point, should be interesting to investigate the homogeneity of the down$-$time databases 
and to check also the down$-$time of the other telescopes at ORM as well as to study the local weather
conditions. This is a key point for the study of the site of future
large telescopes.

\section{Conclusions}
We presented for the first time an analysis of long$-$term
temperature data directly obtained from local
 meteorological towers of TNG and CAMC, at a height of about 2300 m above sea level, far from urban concentration
 and well above the inversion layer.\\
Annual mean values show a similar trend between TNG and CAMC. The linear fit of the 20 years long baseline of
 CAMC data gives an increase of annual mean temperatures of about 1.0$^{\circ}$C/10yrs.\\
  It is interesting to note an oscillation of the values with a
period of about 3$-$4 years that seems to be slightly
 smoothed during the last 10 years. Another evidence is the different behaviour of local minimums and local maximums,
 in fact the minimums increase more rapidly then the maximums.\\
A comparison between NAO Index and the annual mean temperatures shows a correlation of 86\% of significance. In fact, the action of positive NAO Index is like a brake for the increase of temperatures, and like an accelerator for the decrease. Vice
  versa, negative NAO Index acts in opposite mode. Moreover, no correlation is found between NAO Index and number of non-observing nights.\\
We also investigated the influence of the temperature on the
astronomical seeing and we have found that the seeing deteriorates
when the gradient of temperature measured at 2 and 10 m above the
ground is greater than $-0.6^{\circ}$C. This can be explained as
consequence of the lower temperature at 2 m because the higher
temperature at 10 m inhibits the thermal convection below the primary mirror height.

\acknowledgments
The authors acknowledge the CAMC and NOT staff for the availability
of the meteorological data on$-$line and the anonymous reviewer for helpful comments. G. Lombardi thanks also
Ernesto Oliva of TNG for the useful informations and data, Jose L.
Mui\~nos Haro of CAMC for his kindness, support with the data and
the informations, Ricardo Javier C\'ardenas Medinas of NOT for his
help and the patience,
 Edward Graham for the informations and his results from Mazo Airport.

\end{document}